\documentclass[12pt]{article} 
\input epsf.tex 
\def\be{\begin{equation}} \def\ee{\end{equation}}
\def\bi{\begin{itemize}} \def\ei{\end{itemize}}
\def\bea{\begin{eqnarray}} \def\eea{\end{eqnarray}} \def\ba{\begin{array}}
\def\ea{\end{array}} \def\ben{\begin{enumerate}} \def\een{\end{enumerate}}

\newcommand{\eqn}[1]{(\ref{#1})}

\newcommand{\prl}[3]{Phys. Rev. Lett. {\bf#1} ({#2}) {#3}}

\newcommand{\hepth}[1]{{\tt arXiv:{#1}[hep-th]}}
\newcommand{\arxiv}[1]{{\tt arXiv:{#1}[hep-th]}}

\def\ep{\epsilon}

\def\br{\nonumber\\}

\begin{document}
{}~
\hfill \vbox{
\hbox{arXiv:2409.nnnn} 
\hbox{\today}}
\break

\vskip 3.5cm
\centerline{\large \bf
Kaluza-Klein discreteness of
 the entropy: }
\centerline{\large \bf
Symmetrical bath and CFT subsystem}

\vskip 1cm

\vspace*{1cm}

\centerline{\sc  Harvendra Singh }

\vspace*{.5cm}
\centerline{ \it  Theory Division, Saha Institute of Nuclear Physics} 
\centerline{ \it  1/AF Bidhannagar, Kolkata 700064, India}
\vspace*{.25cm}

\centerline{ \it  Homi Bhabha National Institute (HBNI)} 
\centerline{ \it  Anushaktinagar, Mumbai 400094, India}
\vspace*{.25cm}

\vspace*{.5cm}

\vskip.5cm


\centerline{\bf Abstract} \bigskip

We explore the entanglement entropy of CFT systems in contact with 
large bath system, such that the complete system lives on the boundary of
$AdS_{d+1}$ spacetime. We are interested in finding the HEE of
a bath (system-B) in contact with a central subsystem-A.
We assume that the net size of systems A and B  together 
remains fixed while allowing variation in  individual sizes. This
assumption is simply guided by the conservation laws. 
It is found that for large bath size the island entropy term
are important. However other subleading (icebergs) terms 
do also contribute to bath entropy. 
The contributions are generally not separable
from each other and all such
contributions add together to give rise a fixed quantity. 
Further when accounted properly all
such contributions will form  part of higher entropy
branch for the bath. 
Nevertheless the HEE of bath system should be subjected to
 minimality principle. The quantum minimality principle 
$ S_{quantum}[B]=\{S[A], S_{total}+S[A]\}_{min}$,  is
 local in nature and gives rise to the Page curve. It is shown that the
changes in bath entropy do capture Kaluza-Klein discreteness. 
The minimality principle would be applicable 
in finite temperature systems as well.
 
\vfill 
\eject

\baselineskip=16.2pt


\section{Introduction}

 The  AdS/CFT holographic duality  \cite{malda} has 
provided a big insight to our understanding of entanglement in
strongly coupled quantum theories. 
Our focus here is on simple cases of the entanglement between 
two similar type of quantum systems with common interfaces. 
Generally one believes that sharing of 
quantum information between systems 
is guided by unitarity and the locality. 
Under this principle the understanding of the 
formation of gravitational black holes like states 
evolving from a loosely bound pure quantum matter state 
and the subsequent
evaporation process (via Hawking radiation) still remains an 
unsolved puzzle. Though
it is believed that the whole evolution process would  be unitary and all
information inside the black hole interior (behind the horizon)
will be recovered once the black hole fully evaporated. Related to this 
aspect  there is a proposal that the entanglement entropy curve for the
bath radiation should bend when the half Page-time is crossed \cite{page}. 
This certainly is often true when a pure quantum system is divided 
into two smaller subsystems. 
But for mixed states of the hawking radiation, 
or for finite temperature CFTs dual of the AdS-black holes,
it is not that straight forward to obtain the Page curve.  
However, important progress has been made recently in certain  models 
by coupling  holographic CFT to an external radiation system (bath), 
and in some other examples by involving  nonperturbative  techniques such as 
wormholes, riplicas and islands \cite{almheri,replica19}. Some
answers to the difficult questions have been attempted.
\footnote{ Also see a review on information paradox along
different paradigms \cite{raju}
and for  list of related references therein; 
see also  [\cite{susski}-\cite{hashi}].} 
Especially for Karch-Randall braneworld
set up of gravity with a bath system one can find massive 
island formulation in the works
\cite{massiveislands}. Other interesting developments on entanglement
islands are included in \cite{penington}

Particularly the  AMM proposal for generalized entanglement
entropy \cite{almheri} involves the hypothesis of  islandic $(I)$ 
contribution, and by including  gravitational entropy
 of respective island boundary 
$(\partial I)$. According to this proposal 
the  `quantum' entropy of a 2-dimensional radiation bath subsystem $(B)$ 
can be expressed as
\bea\label{ficti1}
 S_{quantum}[B]=
[ ext\{ {Area(\partial I)\over 4G_N} + S[B ~U ~I]\}]_{min} 
\eea
This model uses a hybrid type  `gravity plus gauge theory'
holographic model. (The model is primerily
based on QES proposal of \cite{engelhardt}). It includes contribution
of (gravitational) island entropy to the bath entanglement entropy. 
The islands are usually disconnected
surfaces inside bulk JT gravity. The JT (conformally or nearly AdS) gravity is  a dual theory 
of `dot' like quatum system on the boundary. 
The same dot lives on the boundary of an 
infinte $CFT_2$ bath system. 
In the low energy description the JT gravity is treated as a system
which is in contact with a 2-dim radiation bath over a  flat Minkowski
coordinate patch. So there are both field theoretic ($S[B ~U ~I]$) as well as 
gravitational contributions (${Area(\partial I)\over 4G_N}$) 
present in the formula \eqn{ficti1}. Secondly there is a
 need to pick the lowest contribution 
out of a set of many such possible extremas, which may
 include  entropy contributions of islands and the radiation.
\footnote{In other extensions of the
hybrid models one also includes wormhole contributions, see \cite{replica19}.} 
Although complicated looking, the expression in \eqn{ficti1} 
seemingly reproduces a Page-curve for the bath radiation  entropy 
\cite{almheri}. However AMM proposal ignores contributions of several
subleading terms which we shall altogether pronounce  as '{\it icebergs}' 
terms. Here it should be clear that
the icebergs also are contributions of various disconnected elements
to the bath entropy, similar to the islands entropy.

The important feature of AMM-proposal is that it highlights 
the appearance of island inside bulk  gravity.
The islands  are normally situated outside the black hole horizon. 
These  arise by means of a dynamical principle, 
and usually the islands are   associated with 
the presence of a bath system when in contact with the quantum-dot. 
We highlight in present work  that other
 subleading (disconnected) contributions of `icebergs' 
entropies would have to be properly accounted for in the bath entropy.\footnote{
To explain our terminology, the iceberg entropy is pretty new one. 
The term was first coined by the  author in \cite{hs2022}. 
Essentially it is supposed to include all subleading corrections to the 
bath system entropy beyond the `island' entropy. 
The overall iceberg entropy contribution was found to be negative, but it remains a subleading contribution.} 
Here we will be extending our 2-dimensional proposal \cite{hs2022} to 
the case of higher dimensional CFTs. We provide the
picture that no doubt there would exist island contributions but we must
not ignore (subleading) icebergs contributions to the bath entropy. 
If these are ignored we only end up with  incomplete interpretation of 
the Page curve for quantum entropy.  

We clearly show
 that when  island, icebergs and leading (pure) bath entropy 
once added together  as a series, that gives rise to
 a system dependent
constant contribution to the entropy of bath 
system. These contributions are naturally inseparable 
from each other as they  compensate 
each other perfectly well no matter how large or small their 
individual contributions could be.  
 We have explicitly shown this phenomenon for 
 limiting case of a quantum dot  like  system located
at the interface of symmetrical  CFT bath system \cite{hs2022}. 
 Correspondingly  we find that there are 
gravitational contributions to the bath entropy
arising from island and the subleading icebergs, for  
system-A attached to large bath (subsystem-B). We must note  that the
entanglement entropy of two such systems together $(A \cup B)$ 
can be expanded as 
$$
S_{total}[A \cup B]=S_{pure~bath}+S_{island}+S_{icebergs}\equiv S_l
$$
This expansion can always be done whenever system-B is sufficiently 
large  compared to system-A, e.g. see drawing in lower figure in \eqn{fig22b}.
\begin{figure}[h]
\centerline{\epsfxsize=3in
\epsffile{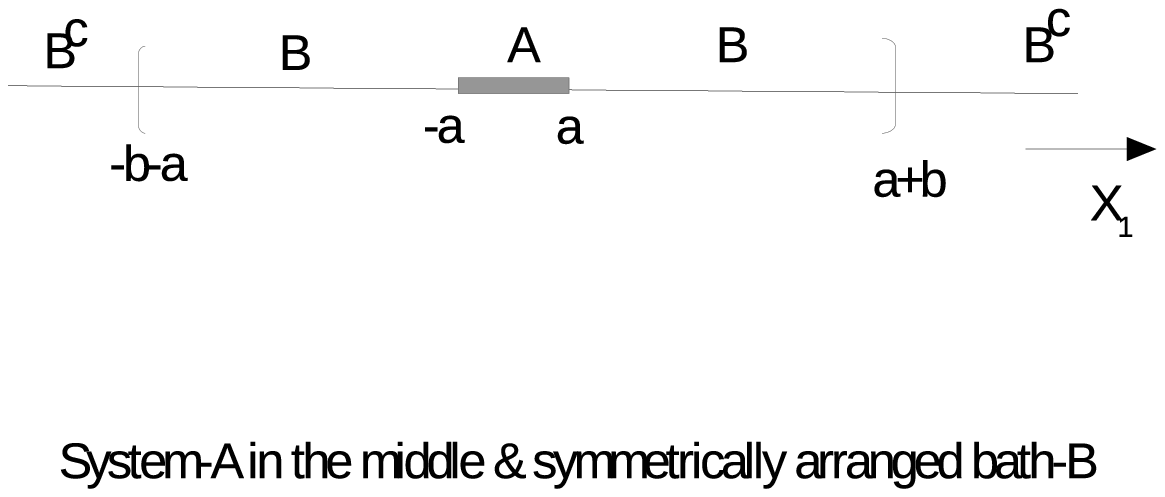} }
\caption{\label{fig23b} 
\small\it A symmetrical arrangement of  CFT system-A in the center (along $x_1$-axis) and 
CFT bath subsystem-B (of size $b$) situated on either side. The transverse spatial directions
 other than $x_1$ (if any) are all suppressed. 
The complimentary system $B^c=(A\cup B)^c$ is also drawn. }
\end{figure} 
  
Note $S_l\gg 0$ is  the  entropy contained in the
total system ($A\cup B$). It is a constant quantity once the total system size is fixed,
and if the system being conserved.
 
Furthermore, it is proposed that 
the quantum entropy of entanglement  of a large bath subsystem-B
should be obtained from local minimality principle
$$ S[B]_{quantum}=\{S[A],~ S_l+S[A]\}_{min}=S[A]   $$
where $S[A]$ is the entropy of system-A. 
To make it further clear the above minimality principle  works because
both $S[A]$ and $S[B]$, as systems being in contact,  involve identical  {\it local} entropies, 
corresponding to common RT surfaces they do share.
\footnote{
 This proposal obviously
excludes all fixed constant values, such as $S_l$, in the entropy of large baths!
Particularly $S_l$ is just a constant and
 remains unaffected under small changes at the common interface
between system-A and bath system-B. We assumed nothing happens
at the interface between system $B$ and $B^c$. 

Secondly, the proposal is valid only when $b\gg a$, i.e.
after  phase transition to the Page curve has taken place. 
There would be further new corrections
to above result, but these will be suppressed in the powers of ratio
$(s\equiv{a\over b})\ll 1$.
We shall report about these corrections in  separate communication.

Regarding the violation of subadditivity: We conclude that the subaddivity
should not apply for quantum entropy $S[B]^{quantum}$ and $S[A]^{quantum}$,
especially in the regime when the bath is very large. In other words, 
when the bath is sufficiently
large compared to system-A

$$S[B]^{quantum} + S[A]^{quantum}=2S[A]<S_l=S[A\cup B]$$

would hold. This is independent of how large the bath may be,
so long as $b\gg a$.} 
Note two systems in contact will have common interfaces fig.\eqn{fig23b}. 
The complementary  system $B^c =(A\cup B)^c$ which is semi-infinite
on both sides does not play direct role here. For non-contact type systems
one should study them separately.
In conclusion, the above two equations reproduce the Page curve 
for the entropy of a large bath system, 
including for finite temperature systems. 
The formula is  definitely valid  at least in  
static (equillibrium) cases. 
For  time dependent  processes involving
black hole evaporation, or if there is a continuous change in the 
total system size (e.g. due to change in mass, energy, or horizon size), 
the scenario is expected to be similar at any point in time,
hence it might be applicable for slow enough processes.

The rest of the article is organized as follows. In section-2 we explain 
the island and icebergs contributions and define the generalized entropy
formulation for pure $AdS$ case. On the boundary we take
a finite size system in contact with a  symmetrical bath.
In section-3 we also discuss a limiting case when system-A becomes very small
and appears as point like
(where we would use Kaluza-Klein scenario) 
and is in contact with large size CFT bath. 
Under Kaluza-Klein perspective
the situation emerges similar to that of $2n$ parallel quantum (strip) systems
 attached to some large bath.
The results for the AdS black holes cases are covered in section-4. 
The  section-5 contains a summary.   

\section{Islands and Icebergs and bath subsystem entropy}

Let us consider a system (A) in contact 
with a bath subsystem (B) in fully symmetrical set up, and 
both having  finite sizes, and 
living on the boundary of the $AdS_{d+1}$ spacetime. 
Thus it is assumed that both systems A and B are made of 
identical field species, i.e. described by same field contents, 
for  simplicity of the problem. 
The pure $AdS_{d+1}$ spacetime geometry is described by
 following line element
\bea\label{ads3}
ds^2={L^2 \over z^2} (- dt^2 + dx_1^2 +\cdots + dx_{d-1}^2+ dz^2)
\eea
where constant $L$ represents a large radius 
of curvature (in string length units) 
of  spacetime. 
The  coordinate ranges lie $-\infty\le (t,~x_i)\le\infty$ 
and  $0\le z \le \infty$, where coordinate $z$ 
represents the holographic range of boundary theory.\footnote{ The Kaluza-Klein
compactification  
over a circle (say $x_{d-1}\simeq x_{d-1}+ 2\pi R$)
 produces a
{\it  conformally } anti-de Sitter  solution 
in lower dimensional gravity. Particularly,
 for $d=2$ case one obtains Jackiw-Teitelboim type 2-dim 
dilatonic background \cite{JT,JT1},
\bea\label{gy67}
&& ds^2_{JT}={L^2 \over z^2} (- dt^2 + dz^2)\br
&& e^{-2 (\phi-\phi_0)}= \sqrt{g_{xx}}={L \over z} 
\eea
where $\phi$ is a 2-dimensional dilaton field of the effective bulk 
gravity theory, all written in the standard convention 
(effective string coupling vanishing near the boundary). The two
 Newton's constants
get related as ${2\pi R \over G_{d+1}}\equiv {1 \over G_d}$, 
with $G_2$ being dimensionless for $AdS_2$.}
\begin{figure}[h]
\centerline{\epsfxsize=3.5in
\epsffile{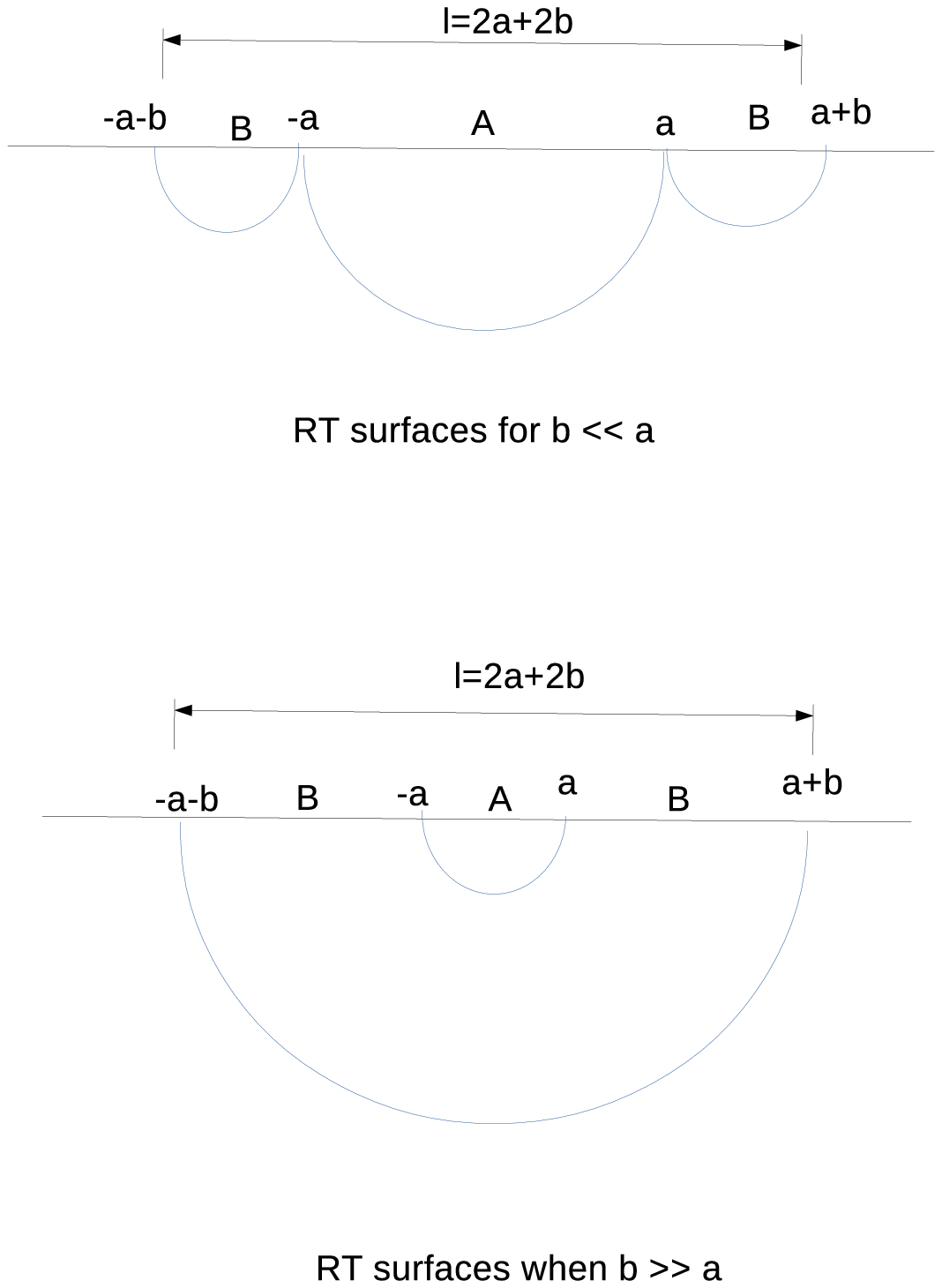} }
\caption{\label{fig22b} 
\small\it 
Two different extremal situations are drawn schemetically. 
The extremal RT surfaces when bath-B size
is  small compared to the system-A ($b\ll a$), we have 
all disconnected minimal area surfaces. 
The lower graph for $b\gg a$ instead has connected surfaces. 
We  keep $l$  same in the both the graphs. 
}
\end{figure} 
 
The $CFT_d$ theory lives on  $d$-dimensional  
$(t,\vec{x})$ flat Minkowski spacetime describing
boundary dynamics of $AdS_{d+1}$ bulk geometry. 
Consider a set up in which a finite size bath subsystem-B 
lives on the coordinate patches
$[-(b+a),-a]$ \& $[a, (b+a)]$ 
along $x_{d-1}$ direction, whereas 
system-A, sandwiched in between the bath, lives over the patch $[-a,a]$. 
The sketches are provided in figures \eqn{fig23b}, \eqn{fig22b} for clarity. 
The entire systems set-up is arranged in a symmetrical way
for the convenience. The states of the system-A and the bath 
subsystem-B are obviously entangled. 
The complementary system $B^c$ lives over the coordinate patches 
$[b+a,\infty]$ and $[-(b+a),-\infty]$. 
The  $B^c$ is semi-infinite on either sides.

From Ryu-Takayanagi holographic prescription  the 
entanglement entropy of a $d$-dimensional $CFT$ strip-shape subsystem
 of width $l~(l=2a+2b)$ is given by 
\bea\label{tot1}
S_{total}[A \cup B]= 
{L^{d-1}V^{(d-2)} \over (2d-4) G_{d+1}}\left(
 {1\over \epsilon^{d-2}}-{b_0^{d-1}\over(a+b)^{d-2}}\right)\equiv S_l
\eea
where  $\epsilon\simeq 0$ is the  UV 
cut-off  of the $CFT_d$ (we shall only consider examples where $d>2$).
$V^{(d-2)}$ is the regulated volume of the box of $(d-2)$ spatial directions
perpedicular to the strip width direction $x_{1}$.\footnote{
Note $b_0={1\over 2(d-1)} B({d\over 2d-2},{1\over 2})$ is specific 
dimension dependent coefficients involving explicit 
Beta-functions, see more in appendix of \cite{hs2015}.} 
We take $l$ sufficiently large but fixed, so that $S_l$  has a constant value.
Our aim is to  determine entanglement entropy 
 where size $b$ is varied from $b\simeq 0$ to $b\simeq l/2$, by hand.  
We simply assume local conservation laws, so that the net gain (loss) 
of system-A is compensated by equal loss (gain) in the size of bath subsystem-B 
and vice-versa. Note that such a process will  keep $l$ fixed.  
{\it Especially for explicit time dependent cases one
 may have some definite rate of change, while  $\dot a=-  \dot b$
is true due to conservation of energy.
The exact rate of loss or gain and the mechanism by which it may happen
is not important here and the actual details of 
the physical process is also not required here.} 
All we are considering is that
local conservation laws are at work for  
complete system within total box size, $l$.\footnote{
Any explicit time dependent processes are not studied here.} 
Obviously we are assuming here that the
system and the bath are made up of identical (CFT) field content.
Let us consider two extreme cases below. 
 
{\it Case-1: Independent entropies} 

When $b\ll a$, for the bath subsystem-B being very small in size, 
the entanglement entropy of the larger system-A  can be found
 by its extremal surface area as \cite{RT,HRT}
\bea\label{fim1}
S[A]
&=& {L^{d-1} V^{(d-2)} \over (2d-4) G_{d+1}}\left( {1\over \epsilon^{d-2}}
-{b_0^{d-1}\over a^{d-2}}\right)\br
\eea
While the extremal surfaces of bath subsytem on both sides become 
disconnected.
The entropy of small subsystem-B become independently, 
\bea\label{fin1}
S[B]= {L^{d-1}V^{(d-2)} \over (d-2) G_{d+1}}\left(
 {1\over \epsilon^{d-2}}-{2^{d-2}b_0^{d-1}\over b^{d-2}}
\right)
\eea
Eq.\eqn{fin1} involves area contributions from two disconnected but identical
extremal surfaces, which contribute
to the bath entropy, see the upper graph as in fig.\eqn{fig22b}.
Note that   
$S[A]$ and $S[B]$ have the local parts which   
depend on individual strip parameters $a$ and $b$, respectively, 
Thus both systems entropies are completely independent even though systems
 are in contact
with each other. The entropy of the small bath system-B would be defined by
  equation \eqn{fin1} until  
the crossover point is reached. After the crossover  new extremal
(connected) surfaces will  emerge as drawn in lower graph of  fig.\eqn{fig22b}.
We will discuss it next.

{\it Case-2: Entropies with identical local components} 

When $b\gg a$, in this regime of large bath system the  entanglement
entropy of system-B is given by the equation
\bea\label{fin2}
S[B]= 
{L^{d-1}V^{(d-2)} \over (2d-4) G_{d+1}}\left( {1\over \epsilon^{d-2}}
-{2^{d-2}b_0^{d-1}\over (l-2b)^{d-2}}
\right)+S_l
\eea
where $S_l$  is total  entropy of the systems together, 
which is a fixed quantity. 
$S_l$ gets the contribution from the outer
 RT surface connecting  two farther ends of the symmetrical  bath on either side, 
follow the lower graph in fig\eqn{fig22b}. 
 Note $S_l$  is independent of individual sizes
 $b$ or $a$, and it
is a fixed quantity for given  $l$. We might still 
 vary  individual system sizes such that we keep $l$ fixed. 
Now for smaller system-A of strip width $2a$ the entropy is given by 
\bea\label{fin2j}
S[A]=
{L^{d-1}V^{(d-2)} \over (2d-4) G_{d+1}}\left( {1\over \epsilon^{d-2}}
-{2^{d-2}b_0^{d-1}\over(l-2b)^{d-2}}\right)
\eea
where we have simply used $2a={l}-2b$.

From  eqs. \eqn{fin2} and \eqn{fin2j}, we observe that
the $S[A]$ and $S[B]$ differ only by an overall constant, $S_l$,
 otherwise they have the same type of {\it local} dependence on $b$.
It also implies that under small change in the system sizes
\bea
&& {\partial S[A]\over \partial b}
={\partial S[B]\over \partial b} \br
 && {\partial S[A]\over \partial a}
={\partial S[B]\over \partial a} \br
 \eea
Put in other words,  $S[A]$ and $S[B]$ 
 represent two independent extrema of the same observable and only differ up to a constant. 
Classically the areas of these extremal surfaces is such that $S[B]>S[A]$, but
 quantum  entropy of  large bath instead
would be governed by the minimization
\bea\label{fin2s}
S^{quantum}_{b\gg a}[B]
=\{S[A], S[A]+S_l\}_{min}=S[A] \ . 
\eea 
This states that the quantum  entropy of a large bath is
the same as the entropy of smaller system-A.
Furthermore we
note that the quantum entropy of bath decreases as  $b$ 
gets larger and larger but at the same rate as that of system-A. 
So the Page-curve for entropy of large bath system ($b\gg a$)
follows from the principle of minimum entropy, 
if there exist multiple
extrema separated by constants, like $S_l$ here. 
The system-A and the bath subsystem-B entropy 
otherwise have identical local dependences. 
This is the net conclusion of 
the proposal given in eq.\eqn{fin2s}. 
Although we might still wonder that the bath entropy
ought to have been  taken simply as $S[B]$ given in \eqn{fin2}, 
which is net classical area of bulk extremal surface. 
Instead, as per quantum minimality proposal \eqn{fin2s}, the  bath entropy
 has to be given by smaller quantity $S[A]$, mainly because both 
have the same local dependence. 
The latter is in agreement with the Page curve
expectation and unitarity for quantum systems that follows from conservation laws.
The quantum entropy proposal shound be taken as  complete result for 
extremal CFTs (at zero temperature) 
being in contact with symmetrical bath. 
 \begin{figure}[h]
\centerline{\epsfxsize=3in
\epsffile{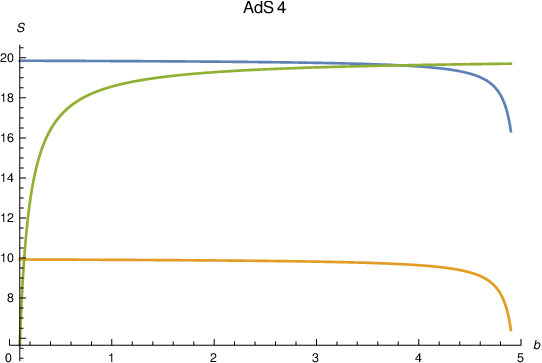} }
\caption{\label{figent3a} 
\small\it 
Entropy plots for the values $l=10,~\ep=.1$, for  $CFT_3$ system. The  
lower falling curve (in yellow) is  preferable for  quantum entropy
of large size bath. The rising (green) curve is  
good for entropy in the small bath
region only. The topmost graph is  classical entropy of large bath and 
 is not the physical one. We have set $ {L^2l_2\over 2 G_4}=1$ for the graphs.}
\end{figure}

\noindent{\bf  Island and Icebergs: full system entropy}

We did not explicitly encounter any island  
 like isolated (disconnected)
surface area contributions to the entropy so far.\footnote{In higher dimensional bulk theories 
the 'islands' are multidimensional (co-dim-2) surfaces. These are no longer 
point objects as in JT gravity and radiation bath models.} 
So where are these contributions hidden in the above analysis? We  have
got the Page curve using quantum minimality principle for large bath system without 
knowing about the  island (fragments) entropy
contributions. To understand it
 we need to dissect the total entropy of  system-A and system-B. 
It is vital to understand first what $S[A \cup B]$ teaches us, even though 
it is only some system parameter and a fixed quantity.
We explore this systematically for $d=3,~4$ and $5$ dimensions. 
(For $d=2$ CFT case, this analysis has been presented in \cite{hs2022}.)

i) For $d=3$ CFT: The  total 
system  entropy $S_l$ is given by ($l=2a+2b$)
\bea
\label{find3ab}
S_l=S[A \cup B] &= &
{L^2 l_2 \over 2 G_{4}}\left( {1\over \epsilon}-{b_0^2\over a+b} \right) 
\eea
where $l_2$ is the length of the strip. Making an expansion
on the r.h.s.  in small ratio $s\ll 1$ $( ~s={a\over b})$,
 one can find
\bea
\label{find3ab1}
S_l
&= &
{L^2l_2 \over 2 G_{4}}
({1\over \epsilon}- {b_0^2\over b}) +{L^2l_2 \over 2 G_{4}}
{b_0^2\over b}(s-s^2+s^3- \cdots )  \br
&\equiv &
S^{(0)} + S_{Island}
+ S_{Icebergs}
\eea
where expressions in the last line have been identified as:
1) entropy of CFT system of size $2b$ (when there is no system-A):
\bea\label{dind3ab1}
&&S^{(0)}={L^2 l_2 \over 2 G_4} 
({1 \over \epsilon}-{b_0^2 \over b})
,~~~ ~~\eea
2) subleading gravitational entropy of {\it island-ic} boundary:  
\bea\label{j7t}
&&S_{island}=
{L^2 l_2 a\over 2 G_4} {b_0^2\over b^2}, ~~~\eea
3) rest all sub-subleading ({\it icebergs}) entropies together as:
\bea
 &&S_{icebergs}=-
{L^2l_2b_0^2 \over 2 G_4} \left( {a^2 \over b^3}- {a^3 \over b^4}
  + \cdots \right). 
\eea
Note $S^{(0)}$  in eq.\eqn{dind3ab1} represents
the HEE of  $CFT_3$ strip system having width $2b$ (net size of the bath 
subsystem). Note it is  entropy of pure
CFT system without the presence of system-A.
Whereas  $S_{island}$, which genuinely represents 
the interactions between bath-B and system-A,  is the 
`gravitational' entropy of island-like boundary situated at $z= b$  
(e.g. corresponding to an 
 island-like region lying in between $z=\infty$ and $z=b$ 
inside the $AdS_4$). 
The codim-2 island boundary  located at $z=b$ has a 
 geometrical area:

 $${\cal{A}}_{island}={2L^2 l_2 a \over b^2} $$ 
It is  proportional
to actual size $2 l_2 a$ of boundary CFT system-A. Thus 
geometric entropy of island eq.\eqn{j7t} may be expressed as
$$ S_{island}= {{\cal{A}}_{island}\over 4\tilde G_4}\equiv
{L l_2 \Phi_0\over 2G_3 b^2}$$
Here $ \tilde G_4\equiv G_4/b_0^2$ and $G_3=\tilde G_4/(2\pi R)$ has been
 treated as effective 3-dimensional  Newton's constant (refer to Appendix).\footnote{
It can be seen if we first dualize the CFT
strip system-A into $AdS_4$, 
and subsequently do $S^1$ compactification 
(along $x_1$ direction and take radius of compactification as $\simeq a$) 
to obtain 3-dimensional `near-$AdS_3$'.
That would create an equivalent
hybrid `near AdS3' plus CFT3 system (as bath) on either sides.
This hybrid gauge-gravity set-up has
been clarified in the Appendix. While 
CFT-near AdS-CFT is only an effective set-up but it
provides a good physical interpretation of the
second term (island) in the expansion of the $S_l$.}
 Other sub-subleading
terms, which are altogether named  as icebergs' entropies, 
include contributions from  remaining  terms
 in  small $s$  expansion.\footnote{
These are called as icebergs (one may even call Archipelagos)
because their total contribution is overall negative definite, 
even if it remains sub-leading \cite{hs2022}.  
The physical interpretation may be difficult to get 
for all these terms.}
 Thus by doing a series expansion it
gets revealed that 
the various terms in the series \eqn{find3ab1} although may have got
distinct interpretations, 
but they  are actually inseparable from each other. 
No matter what these
individual values might be, all the terms are  important
 because these add up nicely to constitute  
   $S_l$, i.e. the total entropy. 
 The total entropy, including island and icebergs, 
thus has a constant value, for given  $l$. (The 
$S_l$ depends only on the parameter $l$, which is measure of total systems size. 
In this sense $S_l$ is actually a global quantity.)
One could still vary $a$ and $b$ individually but
keeping $l(=2a+2b)$ fixed.
That means the bath size could grow  at the cost of size of system-A  
and vice versa,  under  mutual local exchanges or those processes
which may lead to shift in the mutual interface  of 
 systems A and B. 
(This  appears akin to what might happen in black 
hole evoparation processes also, e.g. through Hawking radiation, 
where the Hawking radiation is treated as bath.)
Perhaps simple  CFT models may teach us something 
about the black hole evoparation process! 
The relevant entropy graphs are plotted in the figure \eqn{figent3a}
with a discussion in the caption for small and large bath cases.

{\it Unitarity and the locality of entropy}

The question still arises what would happen if we  
 ignored the subleading (icebergs) contributions in eq.\eqn{find3ab1}, 
due to their smallness, as being subleading  $O(s^2)$ and even smaller. 
Although one is free to do so but we immediately  find that the r.h.s. of 
$S_{l}$ will now start depending on $a$ and $b$
in  independent ways! This will lead to 
 varied conclusions regarding the Page curve, 
 unitarity and about the information content of the systems. 
Alternatively we may decide that  $S_{icebergs}$ terms
 should not be dropped from leading terms in any  situation. 
In other words precise knowledge of isolated contributions is vital for the unitarity!
Furthermore, the island and the icebergs would remain invisible, 
as these contribute to eq. \eqn{fin2}, which is an higher entropy
extrema and hence unphysical as per `quantum 
entropy' proposal of bath-B. The latter reason solely 
arises from  the quantum  entropy principle
that for large size bath entropy should be taken with smaller value
amongst $(S_l + S[A])$ and $S[A]$, where $S_l\gg 0$ is total  entropy. 
 
ii)  $d=4$ case: For $CFT_4$ the total system entropy $S[A\cup B]$ can be written as
\bea\label{find4ab}
S_l &= &
{L^3 l_3 l_2\over 4 G_{5}}\left( {1\over \epsilon^2}-
{b_0^3\over (a+b)^2} \right) 
\eea
where $l_2 \& l_3$ are the sizes of two transverse spatial
coordinates of the $CFT_4$ on the boundary of $AdS_5$ geometry. The systems are separated
along $x_1$ direction.
By making an expansion
on the r.h.s. in \eqn{find4ab}, for  small  $s$
 we get
\bea
S_{l}
&= & {L^3 l_3l_2 \over 4 G_{5}}\left( {1\over \epsilon^2}
-{b_0^3\over b^2} +{b_0^3\over b^2}s (1+s)^{-2} \right) \br
&= &
{L^3 l_3l_2 \over 4 G_{5}}\left( {1\over \epsilon^2}
-{b_0^3\over b^2} +{b_0^3\over b^2}(s-2s^2+O(s^3) ) \right) \br
&\equiv &
S^{(0)} + S_{Island}
+ S_{Icebergs}
\eea
where break up of various expressions in the last line is as:
the leading pure CFT system entropy of size $2b$,
\bea
&&S^{(0)}={L^3 l_3l_2 \over 4 G_5} 
({1 \over \epsilon^2}-{b_0^3 \over b^2})
,~~~ ~~\eea
the  island-ic (gravitational) entropy,
\bea
&&S_{island}=
{L^3 l_3 l_2 b_0^3 a\over 4 G_5 b^3} \equiv{{\cal{A}}_{island} \over 4 \tilde G_5}
, ~~~\eea
and other subleading icebergs entropies:
\bea
 &&S_{icebergs}=-
{L^3 b_0^3 l_3l_2\over 2 G_5} \big[ {a^2 \over b^4}- {a^3 \over 3b^5}
  + \cdots \big]. 
\eea
Note again $S_{island}$ is proportional to  ${2 L^3 l_3 l_2 a\over b^3}$ which
is the geometrical area
of extended 3-dim island boundary located at $z=b$ inside $AdS_5$ spacetime,
with a redefined 5-dim Newton's constant as $\tilde G_5=G_5/b_0^3$.
The related entropies are plotted in the figure \eqn{figent4a}.
They exhibit similar properties as $d=3$ case for small and large bath cases.

iii) For $d=5$ case: 
Above results clearly tell us that for  $CFT_5$  
the subsystem entanglement properties would also be similar to $d=3,~4$ cases. 
We may convince ourselves that it will be a common occuring 
phenomenon whenever there
are two systems (system and the bath) in contact,
i.e. separated by interfaces. The islands and icebergs typically arise as a result of
interactions and information sharing between the dofs of the systems.

\section{\bf Lower dimensional  Kaluza-Klein perspective: 
Multiple thin strips}
\begin{figure}[h]
\centerline{\epsfxsize=3in
\epsffile{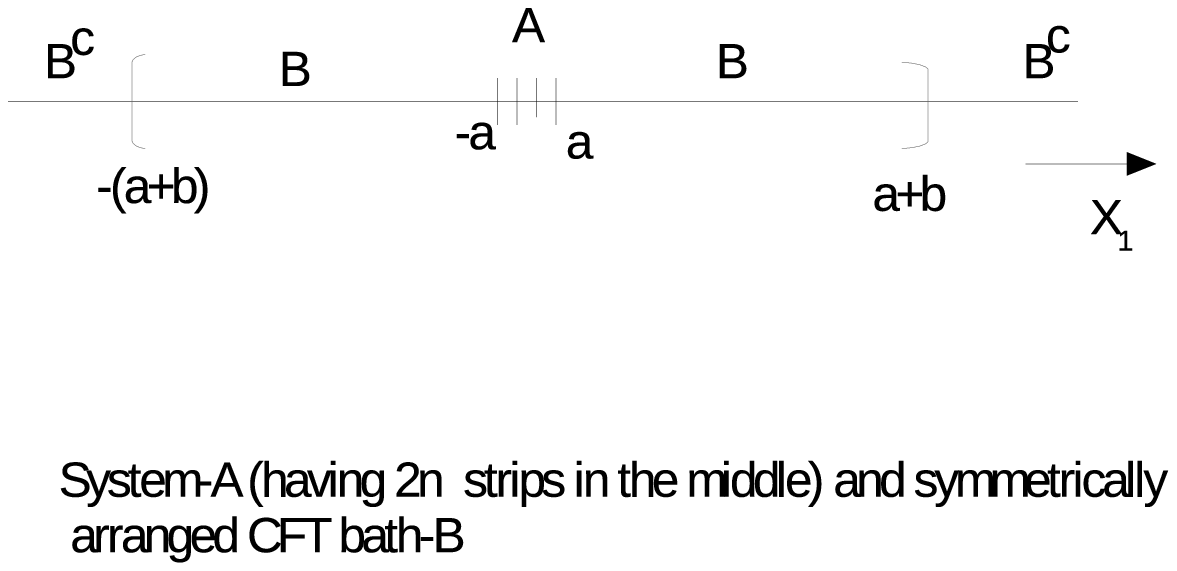} }
\caption{\label{fig21b} 
\small{\it An arrangement of system-A (in the middle on $x_1$-axis) and 
 CFT bath subsystem-B of size $b$ each lie on either side. All transverse spatial directions
(if any) except $x_1$ direction are suppressed. 
The complimentary system $(B^c)$ is also drawn.} }
\end{figure} 
 
It is important to discuss  special case of 
the systems described by  eq.\eqn{fin2}. 
 Consider  very small size for system-A, such that $a\simeq R$, 
where $R$ is  Kaluza-Klein scale of the theory 
(analogous to JT-gravity and near $CFT_1$ case).
(We expect that for such small size system, with a narrow width, system-A 
could be effectively treated as being compactified on $S^1$, with a compactification
radius $R$, but $R\gg\epsilon$.) In that case
we can safely take $a\simeq \pi n R$, with $n$ being 
a positive integer. 
\footnote{Note that it is  only an assumption that there is 
intermediate Kaluza-Klein compactification
on $S^1$ when  at shorter scales the 
size $a$ becomes approximately  $\simeq  R$.
Then it becomes plausible to study a  
lower dimensional dual (gravitational) description 
for system-A (viewed as mutiple strips of narrow width $ \pi R$ 
wrapped along circle (KK direction) but
 fully extended along other transverse directions).}
For simplificity we shall consider only small 
 $n$ values,  also $\epsilon\ll \pi R\ll b$
(If there is any difficulty one can simply take $n=1$). The 
system-A  can essentially be treated as  assembly of $2n$  
narrow (parallel) strips
sandwiched between symmetrical bath system-B (on either sides
of total size $2b$). We have a situation
as depicted in the figure \eqn{fig21b}, 
where transverse directions $x_2$ and $x_3$ are suppressed.  
An expansion
on the r.h.s of \eqn{fin2}, for  small  $ { n \ll {b\over \pi R}}$, 
gives bath entanglement entropy (for $d=3$)
\bea\label{fin2an}
S[B]
&&=S_l  +{L^2 l_2 \over 2 G_4}({1\over\epsilon}-{b_0^2\over n \pi R})\br
&&\equiv ( S^{(0)} + S_{island}+ S_{icebergs})
+S_{2n-strips}[A]
\eea
where leading term $S^{(0)}$ is same as described earlier, and 
other expressions are  
\bea
&&S_{island}= {L^2 l_2\over 4 G_3} {n\over  b^2}, ~~~
\br 
&&S_{icebergs}=-{L^2 l_2\over 4 G_3} 
 ({n^2\pi Rb_0^2 \over  b^3}) + O(n^3) 
\eea
The $G_3=G_4/(2 \pi R)$ is 3-dimensional Newton's constant.
The islandic contribution, especially for $n=1$, 
is similar to the gravitational entropy
of the island boundary situated at $z=b$ inside bulk.
The $S_{icebergs}$ includes rest all subleading contributions. 
The eq.\eqn{fin2an} involves an infinite series which is perturbative. 
Actually  it will not be wise to separate icebergs from 
first two leading terms in \eqn{fin2an} at all!
They  all remain important as the total sum 
of these terms (within the parenthesis) sum up to 
 $l$-dependent entropy $S_l$. 
It is quite clear from the starting line of the 
perturbative expansion in \eqn{fin2an}. 

While the entanglement entropy of the $2n$ parallel
strips, assembled side by side, and in contact with bath, 
 is simply (for $d=3$) 
\bea\label{fin2an2}
 S_{2n-strips}[A]=
{L^2 l_2 \over 2 G_4}({1\over\epsilon}-{b_0^2\over n \pi R})
\eea
and the bath system entropy is
\bea\label{fin2an1}
S[B]=S_l  + S_{2n-strips}[A]
\eea

The local $R$-dependent terms  in eqs. \eqn{fin2an2} 
and \eqn{fin2an1}
are identical, and since the two expressions differ 
only by an overall constant $S_l$,
the actual `quantum  entropy' of bath 
would be given by the local entropy contained in system-A  (multiple
narrow strips) only. So under  minimality selection rule
quantum entropy of bath should simply be stated as
\bea
&&S^{quantum}[B]
=\{S_{2n-strips}, S_l+S_{2n-strips} \}_{min}=
S_{2n-strips}\eea
where entopy of all $2n$ parallel strips arranged together is given above.
Note  $l_2$ is length of these strips, and the individual strip 
width is taken as $\pi R$. 
This is the net entanglement entropy of a large bath system, 
when $b\gg R$, i.e.
towards the end of the Page curve.
It entirely gets its contribution
from single RT surface homologous to  central $2n$ strips system. 
Note the narrow strips have width $ \pi R $ $(\ll b) $.
It is important to note  that
 the bath entropy is  discretized since KK-level $n$ takes integral
values ($n \in Z$). 
Obviously we would trust these results for small $n$ values only. 
For  large $n$ it would be good to use  noncompact continuum 
description in one higher dimensions. 

It can be concluded that the entropy for a large bath will necessarily show 
{\it discrete} jumps as and when KK level changes. This 
is an example of
 strongly coupled system of $2n$ parallel sheets (system-A) 
placed in the middle of large symmetrical
CFT baths on either side. The conclusions shall remain unchanged even under
 infinite size limit $(b\to\infty, ~l\to\infty)$ 
and thus should be treated  universal. 
We conclude that we would not be able to see island 
and icebergs physically,
as their net contribution to entropy always results in a fixed constant only.

\subsection{Entropy spectrum of strip systems}
We are  interested in exchanging  small number of
 strips between bath (B) and the multi-strip system (A).
This will lead to changes in the systems entanglement entropy. 
Note a small number of strip exchanges between B and A will 
not change the total entropy $S_l$.
Thus net change in entropy of system-A
with KK-strips number $n_2$ and with KK-strips number $n_1$ 
can be found to be  $(n_1<n_2)$
\bea
\bigtriangleup S_{1\to 2}&&=S_{strips}[n_2]-S_{strips}[n_1]\br
&&
={L^2 l_2 \over 2 G_4}({1\over n_1}-{1\over n_2}) {b_0^2\over \pi R}\br
&&
\equiv {1\over T_E} \bigtriangleup E_{1\to 2} 
\eea
In the last line the
entanglement temperature of bath can be taken as $T_E \simeq {1\over  l}$ \cite{jyoti}. 
Empirically we may determine that typical change in energy density of systems
(energy per unit strip-length)
\bea
{\bigtriangleup{\cal E}_{1\to 2}} = 
{\bigtriangleup E_{1\to 2}\over l_2} = 
{L^2 b_0^2 \over 2\pi G_4 l R}({1\over n_1}-{1\over n_2}) =
{L^2 b_0^2 \over 4\pi^2 G_3 l R^2}({1\over n_1}-{1\over n_2}) 
\eea
This energy spectrum  is obviously discrete in nature!
A fixed quantum of energy exchanges is required to take place
between bath-B and  multi-strip system-A during an exchange of the strips. 
Note we are discussing the CFT ground state (zero temperature) only.
The spectrum appears analogous to the  atomic 
spectrum, but system-A here is made up of discrete number
of strips  whereas a finite number of strips (quantum matter and not the photons) are  
 exchanged between various KK levels. This exchange
process entails KK-level `jumps'. 
Here $n_1=1$ may be treated as the lowest level 
(for smallest size single strip system)
while  $n_2>1$ will correspond to
 higher levels (more than one strip cases).
Note a level jump in the strip number
is necessarily associated with discrete 
(quantum) CFT matter exchanges between system and its surrounding bath. The 
relevant physical scale is compactification radius $R$. 

A plausible
interpretation of above energy spectrum may be given as follows. The 
discrete KK modes have momentum $\sim {1\over R}$.
(The strip length $l_2$ is some large quantity (fixed), 
$L\gg 1$ is  $AdS$ radius of curvature, 
$G_3\equiv G_4/(2\pi R)$ is 3-dim Newton's constant.) 
Thus $$ \bigtriangleup E_{1\to 2}\propto 
{1\over  R^2}$$ 
 appears primerily due to
the KK momentum modes in this case. 
(The KK-momentum is quantized as $p\sim{n\over R}$, 
where $n\in Z$ is discrete. So $\triangle E\sim p^2$.)  The energy-matter 
exchanges between central strip system-A and the bath 
are precise and discrete! We emphasize that
with more careful
analysis of subleading terms in the entropy, 
we might be able to see winding modes contributions! 
We hope to report about it in subsequent communications.

A similar analysis for 4-dimensional $CFT_4$ would provide  energy density in exchanges
\bea
{\bigtriangleup E_{1\to 2}\over l_3 l_2} = 
{L^3 b_0^3\over 8\pi^3 G_4 l R^3}({1\over n_1^2}-{1\over n_2^2}) 
\eea
where $(l_3l_2)$ is the transverse size of 
the $3d$ strips, where the individual width of strips is $\sim\pi R$.

We will  show that  change in entanglement entropy under 
strip exchange (or KK level jumps) can be determined  for finite temperature cases as well. 
There the $ |\bigtriangleup E|$
will involve thermal corrections. Perturbatively we shall estimate  
and find that
the leading thermal correction grows proportional 
to $R$, the width of the strips. We guess
these corrections involve string winding modes.

\section{Finite temperature systems}

The previous exercise can be extended to the
case of CFT at finite temperature as well. The whole process 
goes in parallel for any CFT system which is in a mixed state. The 
limitation is
that for $d>2$ case the HEE at finite temperature can only be estimated by  
using perturbative methods. Alternative options would be resort to 
 the numerical approach.
Consider the  asymptotically  $AdS_{d+1}$ geometry that 
has Schwarzschild black hole at the center
\bea\label{btz23}
ds^2={L^2 \over z^2}(-f(z) dt^2 +{dz^2 \over f(z) } + dx_1^2+\cdots
+dx_{d-1}^2)
\eea
where $f(z)=(1-{z^d\over z_0^d})$ with
 $z=z_0$ being the location of black hole horizon. 
There is a finite temperature in the field theory on AdS boundary.
Assume  now that the strip shaped system-A  with strip width $2a$
 is taken  in thermal equillibrium with symmetrical bath system-B 
on either sides, so that system-A is located in the middle 
of bath  (net width of bath system being $2b$). 
Other transverse directions of the systems are infinitely extended.
Here both systems $A$ and $B$ have same temperatures.
We  only discuss the case when $a\ll b$, 
because  $a \gg b$ case is rather straight forward.
The entanglement entropy of the
strip like bath system-B on the boundary of \eqn{btz23} can be written as 
\bea\label{ft1}
S[B]= S_{l}({l, z_0})  
+S({2a, z_0})  
=S_{l}({l, z_0})  
+S({l-2b, z_0})  
\eea
In our notation a functional $S(x, z_0)$ 
represents the HEE for strip system having width $x$ 
obtained from area of extremal surfaces
in a  black-brane AdS geometry \cite{hs2015}. 
The  $z_0$ dependences 
indicate there is finite temperature effect 
(horizon dependence).   
Note the first term on the r.h.s. of \eqn{ft1} 
involves  those constants which treat
 system-A and bath-B together as  single entity. It measures
a global information involving $A$ and $B$.
Only  second term  has  the local entropy information 
regarding central system-A like the size and
 location of interface boundaries  
between system-A and bath-B.

Usually for small width $x$  one can estimate $S({x, z_0})$ 
using  perturbative tools as discussed in \cite{hs2015}. It is the 
best method so long as  we have $x\ll z_0$. Also this is  what one would
be requiring most when approaching the end of the Page curve involving
two systems A \& B. 
Typically the perturbative series looks like
\cite{hs2015}
\bea
S({x, z_0})= S_0(x) +S_1(x,z_0)+S_2(x,z_0)+\cdots
\eea
where leading term $S_0(x)$ is the entropy of AdS ground state, 
see eqs.\eqn{fim1} and \eqn{fin1},   while the first order term is given by 
\bea\label{ther1} 
S_1({x, z_0})=
{L^{d-1}V^{(d-2)}\over 16 G_{d+1}}{(d-1)a_1 x^2\over (d+1)b_0^2 z_0^d}
\eea
note it has explicit horizon $(z_0)$ dependence, and same 
is true for second order term
 and so on. (Here $a_1$ and $b_0$ are specific 
dimension dependent coefficients involving explicit 
Beta-functions, see appendix of \cite{hs2015}).

At the same time the entanglement entropy of central  system-A, 
having  width $2a$, is given by
 \bea\label{fin1a}
S[A]= S({2a, z_0})  = S({l-2b, z_0})  
\eea
In second equality we used $l=2a+2b$, as the bath has size $2b$.
Again one can compare that the  expressions in equations
\eqn{ft1} and \eqn{fin1a} differ only by an over all constant
(global) quantity $S_{l}({l, z_0})$, that 
 depends only on full size (system plus symmetrical bath) $l$. 
From the perspective of system-A $S_l$ is some global fixed quantity. 
Therefore we conclude that these two equations represent two different extrema
 of the same entropy observable, primerily because
both entropies contain {\it identical} local terms. 
The local terms arise from  extremal
RT surface that connects common  interfaces of systems  A and B. These 
local terms contain mutual entanglement information which two systems 
share between them.
Further  eqs. \eqn{ft1} and \eqn{fin1a} 
are telling us that they contain identical 
 entanglement entropy (through local exchanges along common interfaces)
for system $A$ and system $B$. Classically they differ only up to  overall constant.
We conclude that `quantum' entanglement entropy measure at finite temperature
for large bath subsystem-B (i.e. $b\gg a$)  
ought to be taken as the smaller value between \eqn{ft1} and \eqn{fin1a}. 
Therefore the quantum entropy of any large thermal bath system (in contact with
 smaller system-A) would be
 \bea\label{fin17}
S_{quantum}[B]
= \{S[A], S_l+ S[A],\cdots \}_{min}=S[A] \ .
\eea

The above result in \eqn{fin17} is consistent with the expectations
of the quantum entropy Page curve when bath is very large. Note this conclusion 
is independent of how large $l$ might  be, so long as $b\gg a$
is obeyed! 
In infinite bath limit $b\to \infty$ ( $ l\to\infty$,  $a=fixed$), 
the constant $S_l$ has a limit:
 $$S_l(l,z_0)\to S_{BH}(z_0)$$ 
where $S_{BH}$ is black hole entropy. So we will get
\bea\label{fin17a}
\lim_{b\to\infty} S_{quantum}[B] 
= \{S[A], S_{BH}+S[A],\cdots\}_{min}\equiv S[A]=S(2a, z_0)  \ .
\eea
The right hand side of the eqs. \eqn{fin17} and \eqn{fin17a} are identical.
Hence for mixed states  irrespective of the overall system
size,  the entropy of a large bath system
is determined by the entropy of smaller system-A, 
under quantum minimality principle. 
$S[A]$  again accounts for the 
smallest entanglement entropy between two contact systems A \& B, 
even in thermal case!
 \begin{figure}[ht]
\centerline{\epsfxsize=3in
\epsffile{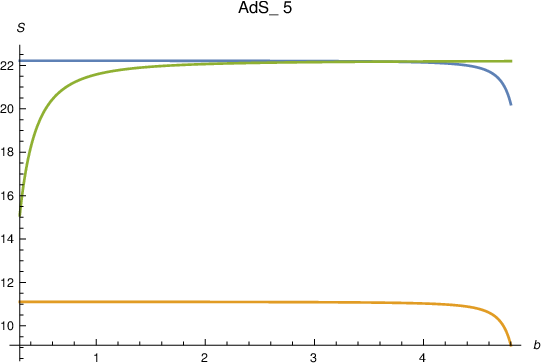} }
\caption{\label{figent4a} 
\small\it 
Entropy plots for the values $l=10,~\ep=.3$, for the $CFT_4$ system. The  
lower falling curve (yellow) is only preferred for  quantum entropy
 at large $b$. The rising (green) curve is  
good only for bath entropy in small $b$
region only. The  topmost graph is  classical entropy for large bath but it 
 is unphysical. We have set $ {L^3l_3l_2\over 4 G_5}=1$.}
\end{figure}

Perhaps an interesting  stage (during blackhole evaporation) can
arise when system-A  size becomes very small such that it becomes
 comparable to intermediate Kaluza-Klein scale of the theory, i.e.
$a\simeq  n\pi  R$, 
and $b\simeq{l\over 2}-\pi n R$, where $R$ is radius of  Kaluza-Klein circle.
Then bath entropy may  be expressed as
\bea\label{fin2am}
S_{quantum}[B]= S({2\pi n R, z_0})  
\eea
which is the entropy of $2n$ independent strips (or sheets)  put together. 
It is the smallest  quantum entropy for a large bath, 
and is quantized by the virtue of the presence of
KK scale in the (bulk) theory. However we should trust 
this result for small $n\in Z$ values
only. It is clear that any change in large bath entropy 
tends to be discrete in nature, if there exists KK scale. 
Similarly, one can deduce that during quantum evolution of a
black hole state after long times, i.e. towards the end of evaporation, 
when system-A will shrink to zero size, the bath entropy
 $$S_{quantum}[B]\to 0\ .$$ That means all the information which 
was kept within system-A
has been transferred to system-B (a very large bath). 
No loss of  information can be expected in this process even for thermal systems! 

\subsection{Spectrum of narrow KK strips}

When the size of the system-A becomes very small such that it is
closer to a Kaluza-Klein radius, $R$, 
we can  set system size as $2a \simeq 2 n \pi R$, for some integer
KK-level $n \ge 1$. ($n=0$ means there is no system-A.)  
We are now interested in exchang of a small number of
   KK strips between bath (B) and the strip system (A). 
We may call this process as `sheets of CFT matter' 
exchange between bath and system-A at their mutual interfaces. This will 
result in change of level $n$ for the system-A.
The resulting change in entanglement entropy  from level
$n_2$ to $n_1$ can be calculated $(n_1<n_2)$ as
\bea
\bigtriangleup S_{1\to2}&&=S_{strips}[n_2]-S_{strips}[n_1]\br
&&
={L^2 l_2 \over 2 G_4}({1\over n_1}-{1\over n_2}) {1\over \pi R}+ \bigtriangleup S_{thermal}\br
&&
\equiv {1\over T_E} \bigtriangleup E_{1\to 2} 
\eea
Assuming 
entanglement temperature can be set as $T_E= {1\over  l}$. So
empirically it can  be determined  that  change in energy 
of strips is  discrete,
\bea
{\bigtriangleup E_{1\to 2}} = 
{L^2 l_2 \over G_3 l (2\pi R)^2 }({1\over n_1}-{1\over n_2}) 
+ {\bigtriangleup E_{1\to 2}^{thermal}}
\eea
A conclusion is drawn from here that a
discrete quantum of energy would have to be exchanged between outer
bath system-B and central (strip-like) system-A for  finite temperature CFT.
 The length of strips $l_2$ is some large value (fixed), 
$L\gg 1$ is AdS radius of curvature,  and
$G_3$ is the Newton's constant. 

The thermal part of entropy is difficult to estimate exactly.
 However in the regime of our interest, when $a \ll z_0$, we can evaluate it perturbatively.
From \eqn{ther1}  up to first order (say, for the $CFT_3$)
\bea\label{ther2} 
 \bigtriangleup E_{1\to 2}^{thermal}
={L^2 l_2\over 32 G_{3} l}{a_1 \pi (n_2^2-n_1^2)R\over 2b_0^2 z_0^3}
\eea  
Thus schematically we get
$$ |\bigtriangleup E_{1\to 2}|\propto {1\over  n_1n_2 R^2}+ \#
{ (n_2+n_1) R\over  z_0^3} $$ 
 Indeed the first  term appears primerily due to
the KK momentum modes. While the second term (due to thermal correction) 
grows linearly with $R$, which we guess
 presumably is  consequence of string winding (wrapping) modes.
We  conclude that these energy and matter exchanges are all discrete! 
The same analysis can also be done for other $CFT_d$.
 
\section{Summary}
We have proposed that quantum entropy of entanglement for 
a large bath system-B (CFT) when in contact with small subsystem-A
 follows quantum minimality principle 
$$ 
S_{quantum}[B]= \{S[A],S[A]+S_l, \cdots\}_{min} =S[A].
$$
where $S_l$ is  full system $(A\cup B)$ entropy.
Any small fluctuation 
in the size of system-A (due to the matter exchange with bath)
would not alter this conclusion provided
systems follow conservation laws.
Thus the equation realizes the Page curve for the entropy of quantum
matter in contact with sufficiently large bath system. This conclusion is based upon the
observation that for small subsystem-A and relatively large bath-B 
the respective entanglement entropies differ only by an overall constant.
The constant does depend on the total systems size $(l)$, 
which is fixed. For this reason $S_l$ is 
conserved and quantum information it contains is essentially global.  
 
We have explicitly shown that islands and subleading entropies (icebergs)
contribute  to the unphysical extremum of bath  entropy. 
Actually all these contributions form various parts of $S_l$ only. 
The (physical) quantum entropy of bath however does not get contribution from 
these fictitious parts! In this light the entropy expression \eqn{ficti1}
is very close to our proposal, but it is only approximate and may not cover full account 
of quantum entanglement between 
systems, mainly it ignores vital subleading contributions 
beyond the  islands. 
On the contrary, we have shown that
there will be an infinitum of such subleading contributions. 
It is shown to be true for all $CFT_d$ systems 
in equillibrium.
Furthermore we have analysed our results 
when subsystem-A size becomes `point-like', similar in size as Kaluza-Klein
scale of the theory, if there exist such an scale. 
As the small system size approaches KK-scale, 
we find necessary discreteness in the entropy
and the energy spectrum due to existence of low lying KK towers. Should 
there be no spontaneous compactification scale
in the theory, the entropy of a large bath would vanish smoothly as and when
 the subsystem-A disappears. In summary it is indicated that the
change in bath entropy does capture Kaluza-Klein discreteness. 
 
\vskip1cm
\noindent{\it Acknowledgments: It is pleasure to thank Stefan Theisen for 
several insightful discussions on this subject. I am thankful to MPI Golm for  
the kind hospitality where part of this work was carried out. The 
financial support from the Alexander-von-Humboldt foundation is
also highly acknowledged. } 

\vskip.5cm
   
\appendix{

\section{An effective construction of the hybrid gravity and CFT systems}
For  small size central CFT subsystem-A, such that its
size $2a\approx 2\pi R$, i.e. when the system size can approximately 
fit within the Kaluza-Klein
radius, the system-A may be treated
as being point like. The symmetrical bath subsystem-B on either side is being comparatively  
very large so it can continue to 
be described by respective (noncompact) CFT. However, for all practical purposes, with out any loss of
physical picture, the 
system-A can also be replaced by dual `near AdS' geometry in one lower spacetime dimensions. 
The Newton's constant for near-AdS bulk geometry
would become $G_{d}=G_{d+1}/(2\pi R)$. We have tried to 
draw these situations in the figure \eqn{figeent23Pen} for systems $A$, 
$B$ and the compliment CFT system $B^c$.
The mixed  gravity and CFT systems set up 
has been a favourable arrangement for an
island proposal \cite{almheri}.

 \begin{figure}[h]
\centerline{\epsfxsize=3.2in
\epsffile{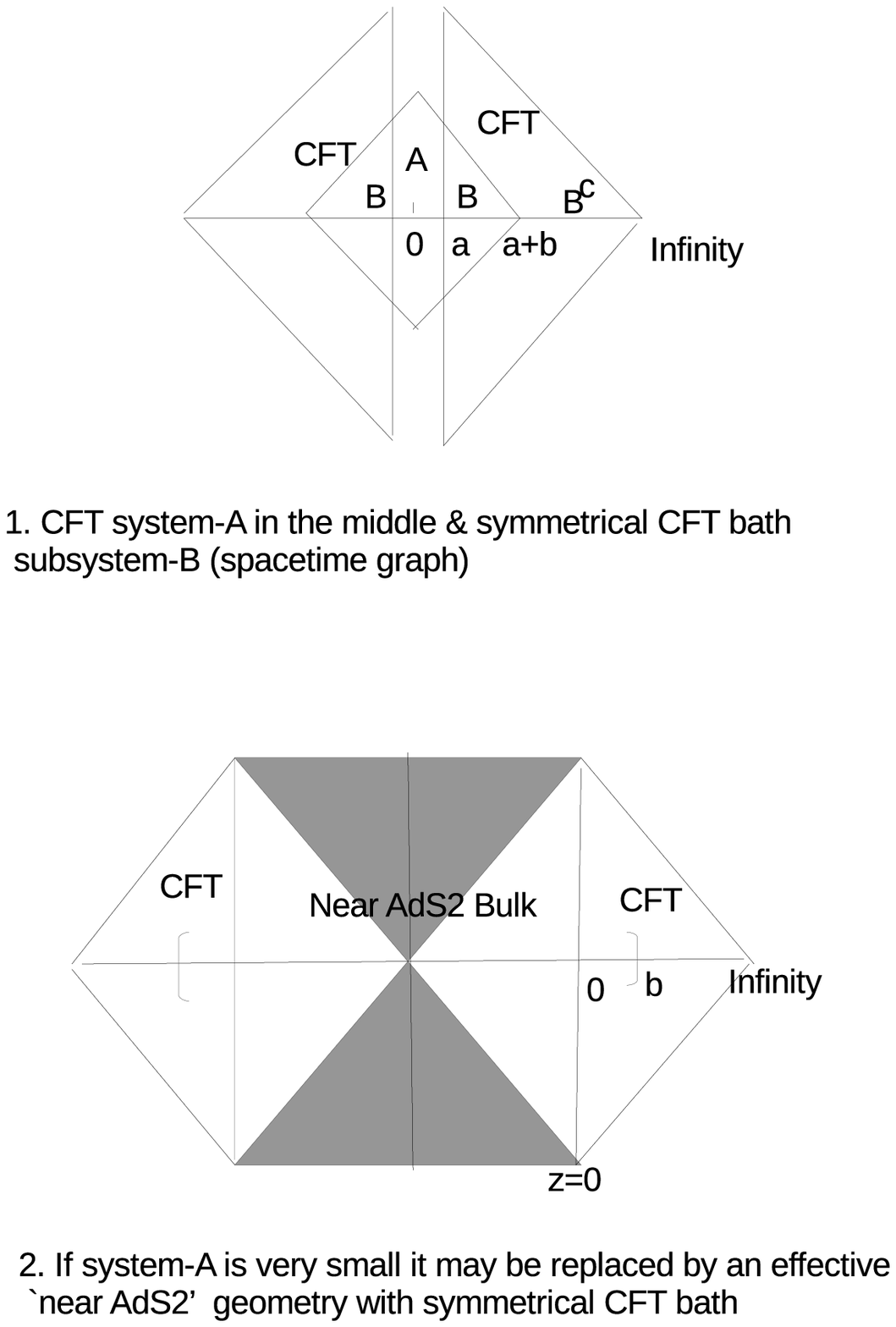} }
\caption{\label{figeent23Pen} 
\small\it An hybrid arrangement of a near $AdS_2$ and $CFT_2$ systems. 
The gravitational constant $G_2$ maybe suitably set. The set up
would remain similar for all  $d$ cases.  }
\end{figure}
}

The entropy of gravitational island located at $z=b$ (symmetrically)
 can be found
\bea
\label{j7t9}
S_{island}={L l_2 \over 2G_3}{\Phi_0 \over b^2}
\eea
for d=3 case. The $\Phi_0$ is suitable
 parameter and is related to
the boundary value of the running dilaton field, $e^{-2\phi}={\Phi_0\over z}$,
 of the near-$AdS_3$ spacetime $ds^2_{(3)}
={L^2\over z^2}(-dt^2+dx_2^2+dz^2)$. 
The islands can be found to be located at $z=b \&-b$, 
and situated symmetrically on either side in figure \eqn{figeent23Pen}. 
Further, without  hassle 
we  set $\Phi_0={L b_0^2 a\over 2\pi R}$, thus eq.\eqn{j7t9}
 would match the eq.\eqn{j7t}. Thus hybrid configuration does provide a
physical understanding of the island term eq.\eqn{j7t}.

\end{document}